\documentclass[twocolumn,pre,superscriptaddress]{revtex4}
\usepackage{graphics}
\usepackage{graphicx}
\usepackage{amsfonts}
\usepackage{textcomp}
\usepackage{amssymb}
\usepackage{mathrsfs}
\usepackage{amsmath}
\usepackage{color}
\usepackage{ulem}
\begin{document}
%\title{Disordered Hyperuniformity Enhances Electronic Transport in 2D Amorphous Silica}
%\title{Topological Transformations in Pentagonal 2D Materials Induced by Stone-Wales Defects}
\title{Topological Transformations in Hyperuniform Pentagonal 2D Materials Induced by Stone-Wales Defects}

\author{Yu Zheng\footnote{These authors contributed equally to this work.}}
\affiliation{Department of Physics, Arizona State University,
Tempe, AZ 85287}
\author{Duyu Chen\footnotemark[1]}
\email[correspondence sent to: ]{duyu@alumni.princeton.edu}
\affiliation{Tepper School of Business, Carnegie Mellon
University, Pittsburgh, PA 15213}
\altaffiliation[present address: ]
{Materials Research Laboratory, University of California, Santa Barbara, California 93106, United States}
\author{Lei Liu\footnotemark[1]}
\affiliation{Materials Science and Engineering, Arizona State
University, Tempe, AZ 85287}
\author{Houlong Zhuang}
\email[correspondence sent to: ]{hzhuang7@asu.edu}
\affiliation{Mechanical and Aerospace Engineering, Arizona State
University, Tempe, AZ 85287}
\author{Yang Jiao}
\email[correspondence sent to: ]{yang.jiao.2@asu.edu}
\affiliation{Materials Science and Engineering, Arizona State
University, Tempe, AZ 85287} \affiliation{Department of Physics,
Arizona State University, Tempe, AZ 85287}

%in photovoltaics, semiconductors, and electronics

\begin{abstract}
We discover two distinct topological pathways through which the pentagonal Cairo tiling (P5), a structural model for single-layer $AB_2$ pyrite materials, respectively transforms into a crystalline rhombus-hexagon (C46) tiling and random rhombus-pentagon-hexagon (R456) tilings, by continuously introducing the Stone-Wales (SW) topological defects. We find these topological transformations are controlled by the orientation correlations among neighboring $B$-$B$ bonds, and exhibit a phenomenological analogy of the (anti)ferromagnetic to paramagnetic transition in two-state Ising systems. Unlike the SW defects in hexagonal 2D materials such as graphene, which cause distortions, the defects in pentagonal 2D materials preserve the shape and symmetry of the fundamental cell of P5 tiling and are associated with a minimal energy cost, making the intermediate R456 tilings realizable metastable states at room temperature. Moreover, the intermediate structures along the two pathways are neither crystals nor quasicrystals, and yet these random tilings preserve hyperuniformity of the P5 or C46 crystal (i.e., the infinite-wavelength normalized density fluctuations are completely suppressed), and can be viewed as 2D analogs of disordered Barlow packings in three dimensions. The resulting 2D materials possess metal-like electronic properties, making them promising candidates for forming Schottky barriers with the semiconducting P5 material.

%The SW defects also preserve hyperuniformity of the materials for all defect concentrations. 

%The P5-C46 transformation embodies a percolation transition of the rhombus-hexagon network, the fraction of which increases linearly as the defect concentration; while the P5-R456 transformation is controlled by the orientational decorrelation of neighboring B-B bonds. 

%However, how the unique DHU disorder affects the electronic and thermal properties in these low-dimensional quantum systems remain elusive. 

\end{abstract}
\maketitle

Two-dimensional (2D) materials such as graphene typically consist of a single layer of atoms packed on a 2D honeycomb lattice \cite{Bh15, Mi14, Xu13} and
possess unique electronic, magnetic and optical properties absent in their bulk form \cite{Bh15, Mi14, Xu13, Yo17}, which enable many novel applications \cite{Bh15, Mi14, Xu13}. Very recently, a class of pentagonal 2D materials have been discovered, which can be derived from the Cairo tessellation with type-II pentagons (see {\bf Fig. 1a}), and realized by single-layer $AB_2$ pyrite structures (see {\bf Fig. 1b}). An example of such 2D material is single-layer PtP$_2$, in which each fundamental cell contains two Pt atoms and four P atoms forming two pairs of pentagons touching through a vertex. Unlike their 2D hexagonal counterparts, 2D pentagonal materials, in particular, PdS$_2$ that have been successfully synthesized exhibit intrinsic in-plane anisotropy useful for various (e.g., thermoelectric) applications \cite{lu2020layer}.

%Mention SW defects, a common topological defects in hexagonal 2D materials, briefly explain what it is, what we knew: preserve hyperuniformity, leads to novel electronic transport properties. Very little is known about SW in Pentagonal 2D materials.

A well-known topological disorder in hexgaonal 2D materials is the Stone-Wales (SW) defect, which can be introduced in the materials via, e.g., proton radiation \cite{St86}. The resulting structures contain ``flipped'' bonds that change the local topology of the original honeycomb network, leading to clusters of two pentagons and two heptagons. The SW defects have been experimentally observed in many 2D materials as local defects \cite{Hu12, Hu13, Ed14, Zh15c, To20} and their effects on the properties of the resulting materials have been numerically investigated \cite{Li11, PhysRevB.86.121408}. Recent studies also suggest that SW defects preserve hyperuniformity \cite{Ch20}, a hidden quasi-long-range order that suppresses infinite-wavelength density fluctuations in the material \cite{To03, Za11b, To16a, To16b}, and could lead to insulator-to-metal transition in 2D amorphous silica \cite{Zh20}.

%However, comprehensive and systematic investigation of the global structures of amorphous 2D materials resulted from these local defects still remain elusive.

%\bigskip
%\noindent {\bf Figure 1:} (a) illustrate the SW defect in PD, emphasize it preseves local shape of the box (b) illustrate the two distinct pathway, highlights two types of bonds with different colors. For pathway, only one types of bonds can flip; for pathway 2, both of them can be randomly flipped. For each path, show an intermediate configuration and final configuration. In all plots, highlights the two types of bonds.
%\bigskip

\begin{figure}[ht]
\includegraphics[width=0.475\textwidth,keepaspectratio]{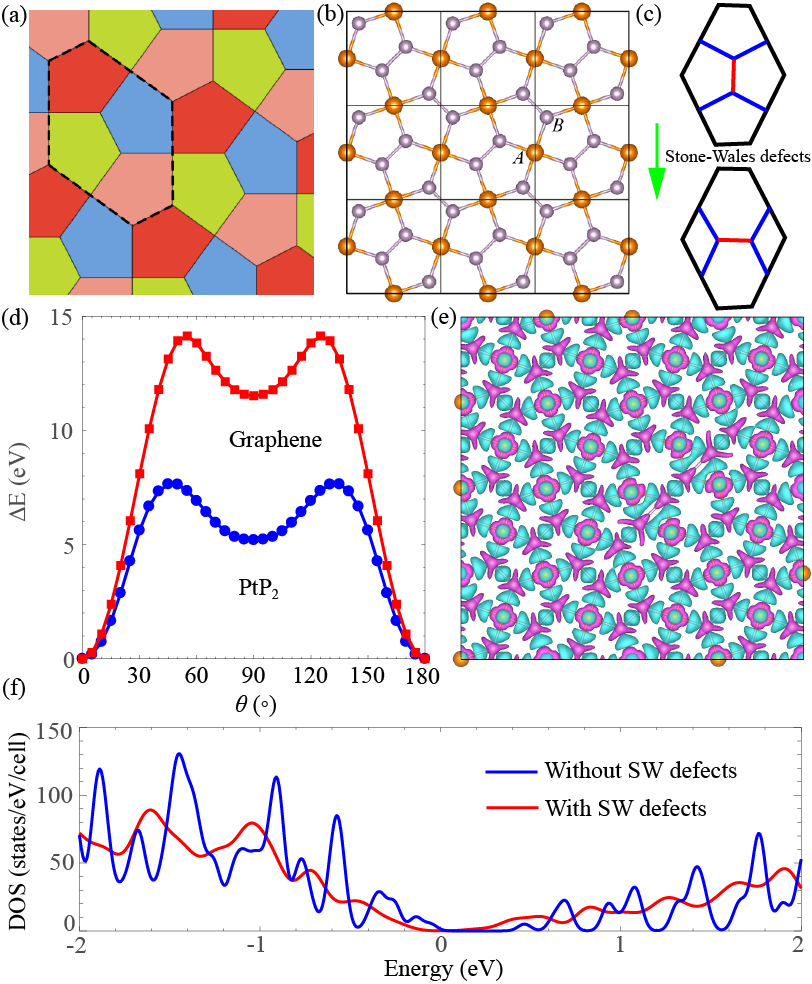}
\caption{(a) Illustration of the Cairo tessellation formed with type II pentagons filling the plane. (b) Illustration of pentagonal 2D $AB_2$ material mapped from the Cairo tessellation. (c) A schematic illustration of the Stone-Wales (SW) topological defect in the Cairo tessellation, which transform two pairs of pentagons in the fundamental cell to a pair of rhombi and a pair of hexagons while maintaining the shape and symmetry of the fundamental cell. (d) Energy difference $\Delta$E of rotating a C-C bond in graphene and of rotating a P-P bond in single-layer PtP$_2$ as a function of rotation angle. (e) Charge density difference of defected single-layer PtP$_2$ with respected to the sum of charge densities of the isolated atoms. The isosurface represents 0.005 $e$ per cubic Bohr radius. Cyan and magenta isosurfaces denote positive and negative values, respectively. (f) Density of states defected and perfect single-layer PtP$_2$.}
\label{fig_1}
\end{figure}

%Talk about introducing SW in PD, refer to Fig. 1a, mention the key results on the structural part, refer to Fig. 1b. Briefly mention the underlying reason, i.e, SW preserve geometry of local unit, and discovery in properties.

Much less is known about the effects of SW defects on the structure and properties of pentagonal 2D materials. We therefore first look into an isolated SW defect and study its energetics and effects on the electronic structure of perfect PtP$_2$. {\bf Fig. 1c} illustrates the local topological transformation induced by a single SW defect in the fundamental cell of the Cairo tilling, which converts two pairs of pentagons into a pair of rhombi and a pair of hexagons touching through a common edge. We perform density functional theory (DFT) calculations (see methods section for detail), where we use a $4\times4\times1$ supercell of single-layer PtP$_2$ and fix the in-plane lattice constants and then rotate the P-P bond (with reference to the center of this bond) from 0 to 180$^\circ$ at an incremental step of 5$^\circ$. In the rotation process, all the atoms are fixed at their locations in an optimized single-layer PtP$_2$ structure (e.g., the P-P bond length is fixed to 2.08~\AA). For comparison, we perform the same energy calculations to obtain the energetics of rotation a C-C bond in graphene. {\bf Fig. 1d} shows the energy difference $\Delta $E as a function of the rotation angle $\theta$. We observe the same trend in the two curves for single-layer PtP$_2$ and graphene, respectively. Namely, a local minimum in the energy difference occurs when the rotation angle is 90$^\circ$, showing the possibility of forming a SW defect in either of the two 2D materials. Furthermore, both curves show two equivalent maxima, one of which is at $\theta$ = 45$^\circ$ and 55$^\circ$ for PtP$_2$ and graphene, respectively. {\bf Fig.1d} also shows that the energy cost to create a SW defect in graphene is much higher than that in single-layer PtP$_2$ owing to the stronger C-C covalent bond. Allowing the atomic coordinates to relax shortens the P-P bond length to 1.98~\AA~and reduces the energy cost associated with this local topological transformation to 2.81 eV. 
{\bf Fig. 1e} displays the charge density difference of a single-layer PtP$_2$ with an isolated SW defect with respect to the sum of charge densities of the isolated atoms. We observe one notable feature, i.e., the charge density distribution at the Pt atoms near the rotated P-P bond no longer has a four-fold symmetry. Instead, two regions of positive charge density difference that are originally separated are merged due to the rotated bond. The isolated SW defect also affects the band gap of single-layer PtP$_2$. Specifically, density of states shown in {\bf Fig. 1f} reveals that the band gap of single-layer PtP$_2$ is reduced from 0.52 \cite{PhysRevMaterials.2.114003} to 0.33 eV. 

%Unlike the SW defects in hexagonal materials, which can induce relatively large local distortions in the honeycomb network, here the shape and symmetry of the fundamental cell of the Cairo tiling is preserved. 

\begin{figure*}[ht]
\includegraphics[width=0.875\textwidth,keepaspectratio]{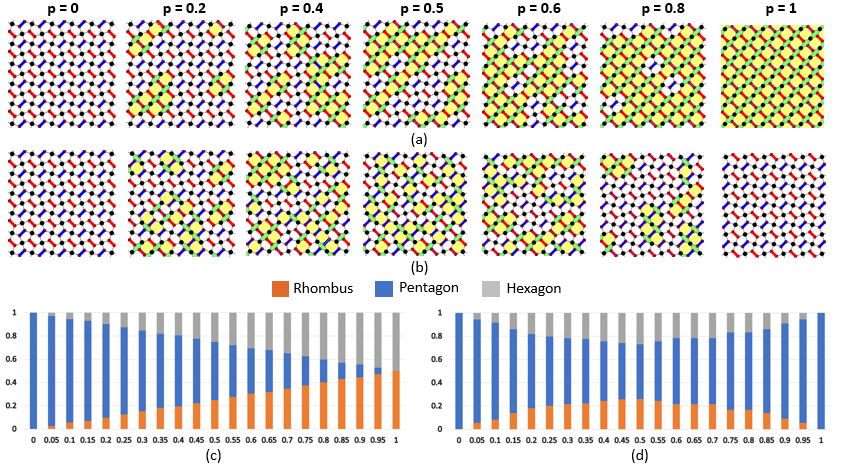}
\caption{Illustration of two distinct topological pathways through which the pentagonal Cairo tiling (P5) respectively transforms into a crystalline rhombus-hexagon (C46) tiling (a) and random rhombus-pentagon-hexagon (R456) tilings (b) by continuously introducing the SW defects. The intermediate configurations along the P5-C46 pathway contain rhombi and hexagons whose orientations of the are all aligned in the same direction. The 4-6 polygon network percolates at the critical defect concentration $p = 0.5$. The intermediate configurations along the P5-R456 pathway contain rhombi (hexagons) that are randomly oriented along two orthogonal directions. The maximal fractions of rhombi and hexagons are achieved at $p = 0.5$. (c) and (d) respectively shows the fractions of the rhombus, pentagon and hexagon as a function of $p$ for the two different topological transformation pathways. }
\end{figure*}

\begin{figure}[ht]
\includegraphics[width=0.495\textwidth,keepaspectratio]{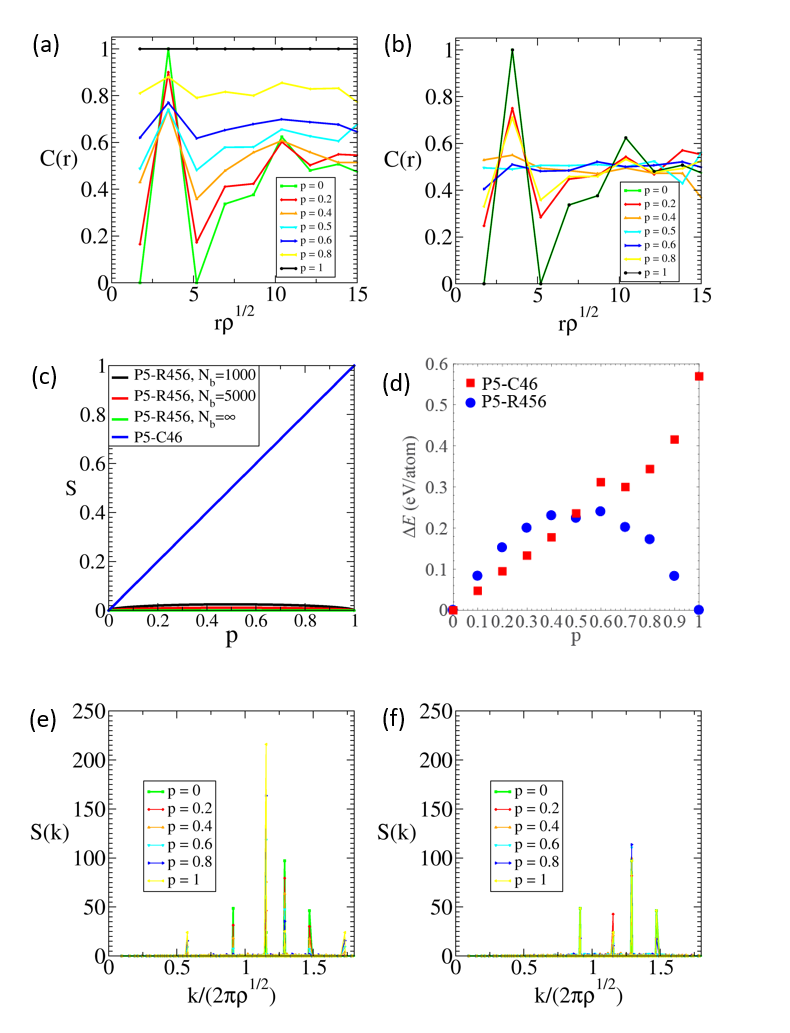}
\caption{P-P bond orientation correlation function $C(r)$ associated with the P5-C46 transformation (a) and P5-R456 transformation (b) as a function of the dimensionless distance $r\rho^{1/2}$, where $\rho$ is the number density of the system.. (c) The global nematic parameter $S$ as a function of $p$ for the two distinct transformation pathways. We note that $S(p)$ associated with the P5-R456 pathway exhibits largest fluctuations around $p = 0.5$ for finite systems. In the infinite-system limit, $S(p)$ converges to exactly 0 for all $p$ as statistical fluctuations vanish. (d) Energy difference $\Delta E$ with reference to the perfect pentagonal structure of 2D PtP$_2$ for the P5-C46 and P5-R456 transition pathways. Structure factor $S(k)$ of the structures along the P5-C46 pathway (e) and P5-R456 pathway (f) as a function of the dimensionless wavenumber $k/(2\pi\rho^{1/2})$, where $\rho$ is the number density of the system.}
\end{figure}

As we will show below, these unique features of SW defects in pentagonal 2D materials lead to novel meta-stable disordered states with distinct topological and physical properties from their counterparts in hexagonal 2D materials. In particular, we discover two distinct topological pathways through which the pentagonal Cairo tiling (P5) respectively transforms into a crystalline rhombus-hexagon (C46) tiling and random rhombus-pentagon-hexagon (R456) tilings as the SW defects are continuously introduced (see {\bf Fig. 2}). These topological transformations are controlled by the orientation correlations among neighboring B-B bonds, and exhibit a close phenomenological analogy of the (anti)ferromagnetic to paramagnetic transition in two-state Ising systems.

%This leads to distinct way how disorder is introduced. As we will show latter, two pathways, etc ... and novel property transition etc...

\section*{Results}

We now describe our numerical procedure for introducing SW defects in an initially perfect P5 tiling and use PtP$_2$ as an example for the corresponding 2D materials. Note that the general methodology discussed here are readily applicable for other 2D pentagonal AB$_2$ materials. As shown in {\bf Fig. 1c}, we apply the flipping operation to the P-P bonds, i.e., by fixing the center of bond and rotation it by $\pi/2$ around the center, to create SW defects. We define the SW defect concentration $p$ as 
\begin{equation}
p = \frac{N_{rot}}{N_{tot}}
\end{equation}
where $N_{tot}$ is the total number of available P-P bonds for rotation and $N_{rot}$ is the number of rotated bonds. Once the topological defects are introduced, we use molecular static simulations to relax the system to a minimal-energy state before collecting structural statistics and perform additional simulations for physical properties. The P-P bonds in the original P5 tiling (and subsequent disordered tilings) assume two perpendicular orientations, which are distinguished with red and blue colors respectively (see {\bf Figs. 2a and 2b}). 

%Mention the tow orientations of the P-P bonds, colored with different colors.

%Mention the methods for introducing SW defects for two different pathways, selection, rotation and relaxation. Emphasize preserving local shape and geometry of repeating unit. Clearly define the defect concetration p 

%restrict to B-B bonds

%\bigskip
%Figure 2: show statistics of 4, 5, 6 as function of p for two different pathways, showing the critical p concentration for the two pathways with dashed red lines, i.e., percolation and maximal fraction; also including the g2 and S_k analysis.
%\bigskip

%he critical defect concentration $p = 0.5$ is illustrated with  concentration for the two pathways with dashed red lines, i.e., percolation and maximal fraction; 

%{\bf Topological transforms induced by SW defects.} 
{\bf P5-C46 Transformation.} We first restrict the rotation operation to P-P bonds with the same orientation. A convenient way to understand this process is to consider that a prescribed fraction $p$ of the fundamental cells containing 4 pentagons of the P5 tiling are randomly selected, and the P-P bonds in the middle of the cells are flipped to create SW defects. This process is repeated for all $p \in [0, 1]$, and the resulting material configurations are shown in {\bf Fig. 2a}. {\bf Fig. 2c} shows the fractions of different types of polygons in the material resulted from the transformation as a function of $p$. Since one can only have three types of polygons, i.e., rhombus, pentagon and hexagon for any $p$ value, the fractions of the three types add up to unity. It can be clearly seen the fractions of rhombi $\phi_4$ and hexagons $\phi_6$ increase linearly with $p$, i.e., $\phi_4 = \phi_6 = 0.5p$; while the fraction of pentagons $\phi_5$ decreases linearly with $p$, i.e., $\phi_5 = 1-(\phi_4+\phi_6) = 1-p$. In addition, the rhombus-hexagon network percolates at p = 0.5 (see {\bf Fig. 2a}). This can be easily obtained by mapping the percolation of fundamental cells containing rhombi and hexagons to the site percolation process on triangular lattice. 

The P5 and C46 crystals at the two extreme concentrations $p = 0$ and $p = 1$ are periodic structures, and are also {\it hyperuniform} \cite{To03, To18a}, i.e., the (normalized) infinite-wavelength density fluctuations in these systems are completely suppressed. This is manifested in their structure factor $S(k)$ as $\lim_{k\rightarrow 0}S(k) = 0$, as shown in Fig. 3(e). Here $S(k)$ is defined as 
\begin{equation}
S(k) = 1 + \rho\Tilde{h}(k),
\end{equation}
where $\Tilde{h}(k)$ is the Fourier transform of the total correlation function $h(r) = g_2(r) - 1$, and $g_2(r)$ is the pair correlation function, $\rho$ is the number density of the system. Note that this definition implies that the forward scattering contribution to the diffraction pattern is omitted \cite{To03, To18a}. Moreover, the P5 and C46 crystals satisfy a stronger condition called {\it stealthy hyperuniformity} \cite{To03, To15, To18a}, for which $S(k) = 0$ for a range of wavenumbers $0\leq k < K$, where $K$ in the crystalline cases is the location of the first Bragg peak. 

The intermediate configurations are random tilings consisting of two types of building blocks of varying concentrations, with each type corresponding to a particular orientation of the P-P bonds in the middle of the fundamental cell. They are neither crystals or quasicrystals, but since the SW defects do not induce any distortion of the fundamental cell, and the number density is strictly preserved on the fundamental cell level, we expect that hyperuniformity in the P5 or C46 crystal is preserved in these random tilings. This is verified by the computed $S(k)$ of these structures as $S(k)$ continues to decrease to extremely small values as $k$ decreases, as shown in \textbf{Fig. 3(e)}. However, we note that the stealthiness of the P5 or C46 crystal is lost in these intermediates structures, i.e., $S(k)$ is not strictly zero at small, yet finite wavenumbers. Nonetheless, the topological transformation in pentagonal 2D materials is a novel class of hyperuniformity-preserving operations that have not been identified before \cite{Ch20, Ki18}. The locations of the Bragg peaks also shift as $p$ varies, reflecting the increasing fractions of rhombi and hexagons in the system. Interestingly, the intermediate structures are tiling variants of the P5 or C46 crystal within the 2D plane, and in this regard can be viewed as 2D analogs of disordered Barlow packings \cite{Mi19}, which are stacking variants of fcc or hcp crystals in three dimensions. 

%Readers can refer to Ref. \cite{Mi19} and references therein for detailed discussion of Barlow packings.} 

We note that transformation is a phenomenological analogy of anti-ferromagnetic to ferromagnetic transition in the two-state Ising system, with the two distinct orientations of P-P bonds respectively corresponding to the two spin polarization directions. The global nematic order parameter $S$ associated with the P-P bonds, which is the analog of the net magnetization for Ising systems, is simply given by
\begin{equation}
S = \frac{1}{N_b}<|n_\parallel - n_\perp|>
\end{equation}
where $n_\parallel$ and $n_\perp$ are the number of P-P bonds along two orthogonal directions, and $N_b$ is the total number of P-P bonds in the system. It is clear that $S(p) = p$ for this transformation, which is also verified from the numerical data shown in {\bf Fig. 3c}. {\bf Fig. 3a} shows the P-P bond orientation correlation function, i.e.,
\begin{equation}
C(|{\bf r}|) = <{\bf s({\bf x})}\cdot{\bf s({\bf x}+{\bf r})}>  
\end{equation}
where ${\bf s}({\bf x})$ is the unitary direction vector of a P-P bond located at ${\bf x}$. The strong oscillations in $C(r)$ with $p=0$ indicates the periodic alternation of bond orientations in the P5 structure, while $C(r) = 1$ with $p = 1$ indicates perfect alignment of all bonds in the C46 structure. The gradual building up of long-range orientation correlation as $p$ increases can be clearly seen from the evolution of $C(r)$.

{\bf P5-R456 Transformation.} We now allow all of the P-P bonds to rotate, but still require that each bond can only be rotated once. The fractions of different polygons as a function of $p$ are shown in {\bf Fig. 2d} and the representative configurations associated with different $p$ are shown in {\bf Fig. 2b}. Similar to the P5-C46 transformation, the initial increase of $p$ from 0 leads to an rapid increase of $\phi_4$ and $\phi_6$ and decrease of $\phi_5$. In the low-$p$ limit, the defects are well separated and independent from one another, and thus, the same linear relations between $\phi_n$ ($n = 4,5,6$) and $p$ as in the P5-R456 transformation also hold. As $p$ further increases, the neighboring perpendicular bond of a rotated bond might also be rotated, which transforms a rhombus/hexagon back to a pentagon. This leads to slower growth of $\phi_4$ and $\phi_6$ (and equivalently, slower decay of $\phi_5$). Therefore, the P5-R456 transform can be considered to be mainly controlled by the orientational decorrelation of the neighboring P-P bonds and the maximal $\phi_4$ and $\phi_6$ are achieved when the ``most decorrelated'' state is reached at $p_c = 0.5$, i.e., 50\% of the bonds are randomly flipped (see {\bf Fig. 2b}). This is the analog of the anti-ferromagnetic to paramagnetic transition in Ising systems. As $p$ further increases beyond 0.5, more rhombi and hexagons are converted back to pentagons, leading to deceasing $\phi_4$ and $\phi_6$. Finally, at $p = 1$, all of the P-P bonds have been rotated exactly once and the tiling transforms back to P5, albeit the entire system is rotated by 90 degree with respect to the original P5 tiling. 

The aforementioned process is also manifested in $S$ and $C(r)$, respectively shown in {\bf Figs. 3c and 3b}. Since random bond flipping on average preserves the number of bonds with either orientation, $S(p)$ remains close to 0 for all $p \in (0, 1)$, with largest fluctuations occurring around $p = 0.5$ for finite systems. In the infinite-system limit, $S(p)$ converges to exactly 0 for all $p$ as statistical fluctuations vanish. On the other hand, the strong oscillations in $C(r)$ associated with the P5 structure are clearly diminished as $p$ increases due to the induced disorder or decorrelation of P-P bond orientations. The most random state characterized by an almost flat $C(r) \approx 0.5$ (i.e., due to equal amount of parallel and perpendicular bond pairs at all separation distances) is achieved at $p = 0.5$. Further increasing $p$ gradually removes disorder, leading to perfect P5 at $p = 1$. 

Similar to the P5-C46 pathway, the intermediate structures in the P5-R456 pathway are neither crystals nor quasicrystals, and yet they preserve hyperuniformity of the P5 crystal. Specifically, their corresponding $S(k)$ approaches zero as $k$ approaches 0, as shown in \textbf{Fig. 3(f)}. The structure factor also shows diminished and shifted peaks as $p$ increase from 0 to 0.5, a common signature of increasing disorder. Moreover, these intermediate structures can be viewed as random tiling variants of P5 (or C46) crystal consisting of finite types of building blocks, with each type corresponding to a particular combination of orientations of the P-P bonds in the fundamental cell.

%The pair statistics $g_2$ and $S(k)$ of the structures for different $p$ values are shown in Supporting Information. Both quantities show diminished and shifted peaks as $p$ increase from 0 to 0.5, a signature of increasing disorder. 

%compared to the intermediate structures in P5-C46 transformation, as the orientation of the rhombi and hexagons can randomly align with two perpendicular directions in the current transformation. As $p$ further increases beyond 0.5, $g_2$ and $S(k)$ evolves towards those of perfect P5 tiling. Moreover, the small $k$ behavior of $S(k)$ indicates the hyperuniformity of all structures (see SI for details), as the rotation of bonds again preserves local number density in the transformation.

%Talk about the second pathway, emaphszie the difference from the first. Refer to the statistics, point out the saturation, i.e., when the orientation of the bonds are relatively random, the number of 4-6 would be stable. Then mention g2 and $S(k)$, evoke similar arguments to convince DHU. Show snapshots of intermediate configurations in SI. [The P5-R456 transition is controlled by orientional decorrelation of the neighboring Pt-Pt bonds and the saturation is achieved when the complete decorrelated state is reach at $p_c = XXXX$ (this can be obtained analytically with very simple arugments), when the half of the bonds are randomly flipped (double check) so neighboring bonds are not correlated in orientation.] Show the evolution of the configuration in the SI.

Figure 3(d) shows the energies $\Delta E$ required to generate SW defects of different $p$ values for the two transition pathways. We can see that, for the P5-R456 pathway, the required energy gradually increases with increasing $p$, which reach the maximum near $p = 0.5$ and decreases with further increase in $p$. On the other hand, $\Delta E$ associated with the P5-C46 pathway increases almost linearly with $p \in [0, 1]$. These trends of $\Delta E$ are completely consistent with the observed structural changes during the two types of topological transformations. We note that the largest $\Delta E$ is much less than 1.0 eV per atom for the full spectrum of structures obtainable from the transformation. 

\begin{figure}[ht]
\includegraphics[width=0.5\textwidth,keepaspectratio]{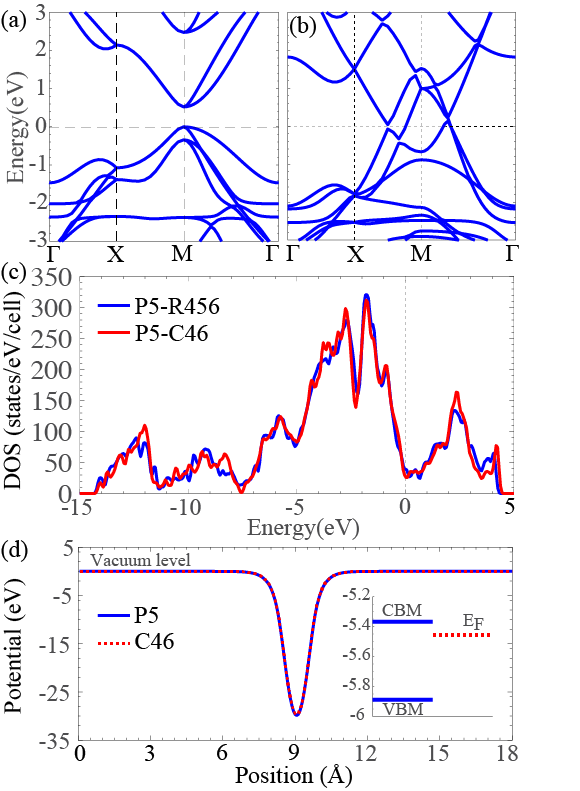}
\caption{Band structures of 2D PtP$_2$ with the (a) P5 (Reproduced from Ref. \cite{PhysRevMaterials.2.114003}  and (b) C46 structures computed at the HSE06 level of theory. (c) Density of states of two intermediate structures ($p$ = 0.5) in the P5-C46 and P5-R456 transition pathways calculated at the PBE level of theory. (d) Potential profile along the slab direction of the P4 and C46 structures. The inset shows absolute energy scales of the conduction band minimum (CBM) and valence band maximum (VBM) of the P5 structure and the Fermi level E$_\mathrm{F}$ of the C46 structure.}
\end{figure}

{\bf Fig. 4a} and {\bf 4b} compare the band structures of the 2D materials derived from the P5 and C46 structures. As reported in Ref. \cite{PhysRevMaterials.2.114003}, the P5 structure is semiconducting with a direct band gap of 0.52 eV. By contrast, the C46 structure is metallic with {\it Dirac-cone-like} dispersion in the conduction bands near the X point. This dispersion is caused by the presence of the hexagonal six-membered rings in the C46 structure. The metallic nature is also shared by the intermediate structures with p values lying between 0 and 1. For example, {\bf Fig. 4c} shows the density of states of the two PtP$_2$ structures with the same p (p = 0.5) from the two different transformation pathways. Both density of states show sizable magnitudes at the Fermi level and their overall shapes are similar in the entire energy window, consistent with their close energies at $p$ = 0.5 as shown in {\bf Fig. 4d}. 

Due to the metallic properties of the C46 structure, we envision a potential application of forming P5-C46 heterostructure as a Schottky barrier. {\bf Fig. 4d} illustrates the average local potentials a carrier experiences in the out-of-plane direction of single-layer P5 and C46 structures. We can see that the quantum wells are nearly identical in spite of the two different atomic structures. We extract the conduction band minimum (CBM; -5.37 eV) and valence band maximum (VBM; -5.89 eV) energy levels of the P5 structure and the Fermi level (i.e. -5.46 eV or 5.46 eV as the work function) of the C46 structure by aligning these three energy levels with the vacuum level set to zero for the two structures. We find that the Fermi level of the C46 structure lies between the CBM and VBM of the P5 structure with the distances from the CBM and VBM being 0.09 and 0.43 eV, respectively.

\section*{Conclusions and Discussion}

In summary, we have studied the effects of the Stone-Wales defects on the structure and electronic properties of a novel class of single-layer $AB_2$ pyrite materials derived from the pentagonal Cairo tiling. Unlike the SW defects in hexagonal 2D materials which cause distortions, the defects in pentagonal 2D materials preserve the shape and symmetry of the fundamental cell of P5 tiling and are associated with a minimal energy cost. We discovered two distinct transformation pathways encompassing a rich spectrum of disordered 2D materials composed of a mixture of rhombus, pentagon and hexagon rings, including the ordered P5 structure and C46 structure as the two extremes. The intermediate structures along the two pathways are neither crystals nor quasicrystals, and yet they preserve hyperuniformity of the P5 or C46 tiling. These random tilings can be viewed as 2D analogs of disordered Barlow packings in three dimensions.

The resulting 2D materials possess metallic properties, making them promising candidates for forming Schottky barriers with the semiconducting P5 material. The semiconductor-to-metal transition in these pentagonal 2D materials as disorder is introduced into the system are similar to the insulator-to-metal transition observed in disordered hyperuniform silica \cite{Zh20}, and suggest a potential broad linkage between disordered hyperuniformity (DHU) and enhanced electron transport in low-dimensional quantum materials, which is complementary to the conventional wisdom of landmark ``Anderson localization'' \cite{An58} that disorder generally diminishes transport. In future works, we will continue to explore this novel concept of disordered hyperuniform quantum materials (DHQM), i.e., disordered hyperuniform atomic-scale low-dimensional materials with nontrivial quantum characteristics \cite{Zh20, Ch20}), which could potentially have broad implications in many fields of condensed-matter physics and materials science.

%Our study also sheds light  

%behavior of a SW defect in 2D pentagonal materials using PtP$_2$ as an example. {\bf To be added by Yang...}

%\bigskip
%\noindent {\bf Figure 4:} Plots for properties: overall energy of the systems with different p; DOS and band gaps calculations etc.
%\bigskip
%{\bf Novel physical properties of transformed structures.}

%Describe the physical property results.

%Mention possible applications.

%End with a brief conclusion and discussion.

%Band structures for both perfect P5 and C46

%DOS plot for the two intermediate states at p = 0.5

\begin{acknowledgments}
H.Z. thanks the start-up funds from ASU. This research used computational resources of the Agave Research Computer Cluster of ASU and the Texas Advanced Computing Center under Contract No. TG-DMR170070. Y. J. thanks the generous support from ASU during his sabbatical leave.
\end{acknowledgments}

\section*{Methods}
\subsection*{DFT calculations}
Density functional theory calculations are performed using the Vienna Ab initio Simulation Package \cite{KRESSE199615} with the planewave cutoff energy of 550 eV, the Perdew–Burke-Ernzerhof (PBE) functional \cite{PhysRevLett.77.3865} and the potentials from the projector augmented wavefunction method \cite{PhysRevB.50.17953}. A $12\times12\times1$ k-point grid and surface slab models with a length of 18 angstroms in the $z$ direction are used in all calculations.  To simulate an isolated Stone-Wales (SW) defect in PtP$_2$ and its two transition pathways, we use $4\times4\times1$ and $6\times6\times1$ supercells,respectively. $6\times6\times1$ supercells of graphene are used to model its SW defect  The in-plane lattice constants of these supercells are fixed to the multiple times (4 or 6) of the in-plane lattice constant of an optimized unit cell of single-layer PtP$_2$. To obtain the density of states (see Fig.1f) with the HSE06 functional \cite{hse06}, we reduce the plane-wave cutoff energy to 400 eV.

%\bibliography{network}

\end{document}